\documentclass[aps,preprint,showpacs,preprintnumbers,amsmath,amssymb]{revtex4}
\usepackage{amsmath,mathrsfs,amsbsy,color,graphicx,bm,amsthm,amsfonts}
\usepackage{units}
\usepackage{bbm}
\usepackage{times}
\usepackage{dcolumn}
\usepackage{mathrsfs}
\usepackage{amsmath,amssymb,epsfig}
\usepackage{amsmath}
\newcommand{\udots}{\mathinner{\mskip1mu\raise1pt\vbox{\kern7pt\hbox{.}}
\mskip2mu\raise4pt\hbox{.}\mskip2mu\raise7pt\hbox{.}\mskip1mu}}
\begin{document}

\title{Quantum coherence and entanglement of mixed states in relativistic settings}
\author{Xing-Le Zhao, Zhen Yang\footnote{Email:yangzhen@lnnu.edu.cn}, Shu-Min Wu\footnote{Email: smwu@lnnu.edu.cn}}
\affiliation{Department of Physics, Liaoning Normal University, Dalian 116029, China}


\begin{abstract}
States prepared in laboratory experiments are generally mixed because of imperfect preparation and residual coupling to the environment, whereas pure initial states represent an idealized limit. Motivated by this distinction, we investigate quantum entanglement and coherence in a pair of Unruh-DeWitt detectors initially prepared in a general two-qubit X state. One detector remains inertial and uncoupled, while the other undergoes uniform acceleration and interacts locally with a massless scalar field in the Minkowski vacuum. We obtain the reduced detector state within a perturbative treatment and evaluate its concurrence, $l_1$ norm of coherence, and relative entropy of coherence. For the initial-state families considered, the resulting expressions exhibit entanglement sudden death at finite acceleration, with larger initial entanglement associated with a higher threshold. Both coherence measures remain nonzero beyond this threshold and approach zero in the formal infinite-acceleration limit of these expressions. We also examine the dependence of these resources on the detector energy gap and interaction duration. The results distinguish the persistence of coherence from that of entanglement in the mixed-state detector model and identify initial mixedness as a relevant ingredient in relativistic quantum information.
\end{abstract}

\vspace*{0.5cm}
\maketitle
\section{Introduction}

Quantum entanglement is a fundamental property of composite quantum states that cannot be expressed as convex mixtures of product states \cite{zxl1}. It is a key resource for quantum information processing, enabling protocols such as quantum key distribution, teleportation, and dense coding to achieve capabilities beyond their classical counterparts \cite{zxl2,zxl3,zxl4,zxl5,zxl6,zxl7,zxl8}. Quantum coherence, which describes superposition relative to a specified reference basis, is another fundamental quantum resource. It underlies interference phenomena and plays an essential role in quantum information processing \cite{zxl9,zxl10,zxl11}. Within the resource theory of coherence, the relative entropy of coherence and the $l_1$ norm of coherence provide computable measures of this resource \cite{zxl12}. Although coherence and entanglement characterize different aspects of quantum states, their resource-theoretic descriptions are closely related. In particular, coherence can be converted into entanglement through suitable incoherent operations on a system and an initially incoherent ancilla, and their relationship has been explored through resource conversion and the distribution of coherence in composite systems \cite{zxl13,zxl14,zxl15,zxl16,zxl17}. Conversion relations between coherence and multipartite entanglement have also been established in frameworks involving incoherent or block-incoherent operations \cite{zxl18,zxl19}. These connections motivate a comparative study of how coherence and entanglement respond to noise and relativistic effects.

At the intersection of relativity, quantum theory, and information theory, relativistic quantum information investigates how motion, acceleration, gravitation, and spacetime horizons affect quantum resources and information-processing tasks \cite{SDF1,SDF2,SDF3,SDF4,SDF5,SDF6,SDF7,SDF8,SDF9,SDF10,SDF11,SDF12,SDF13,SDF14,SDF15,SDF16,SDF17,SDF18,SDF19,SDF20,SDF21,SDF22,SDF23,SDF24,SDF25,SDF26,SDF27,SDF28,SDF29,SDF30,SDF31,SDF32,SDF33,SDF34,SDF35,SDF36,SDF37,SDF38,SDF39,SDF40,SDF41,SDF42,SDF43,SD1,QGL1,QGL2,SD2,QGL3,QGL4,SD3,QGL5,QGL6,QGL7,QGL8,QGL9,QGL10,SDF45,SDF46,SDF47,SDF48,SDF49,SDF50, SDF51,SDF52}. Within this framework, the close connection between entanglement and coherence raises the question of whether these resources exhibit different degrees of robustness under relativistic motion. The Unruh-DeWitt detector model provides an operational setting for addressing this question by describing localized two-level systems coupled to a quantum field along prescribed trajectories \cite{QVMP1,QVMP2,QVMP3,QVMP4}. Uniform acceleration modifies the detector response through the Unruh effect and can alter the quantum resources shared by the detectors \cite{SDF53,SDF54,SDF55,SDF56,SDF57,SDF58,SDF59,SDF60}. Although pure initial states offer a useful starting point, they represent an idealized preparation limit. States prepared in laboratory experiments generally possess some degree of mixedness owing to preparation errors, finite-temperature effects, and residual coupling to the environment. Accounting for initial mixedness is therefore necessary for relating relativistic resource dynamics to realistic state preparations. It also introduces a distinct physical question: entanglement depends on the interplay between populations and off-diagonal correlations, whereas coherence is determined by off-diagonal elements in a specified reference basis. Consequently, their relative persistence under acceleration cannot, in general, be inferred from pure-state results alone. Two-qubit X states provide a tractable setting for this comparison, encompassing both entangled and separable mixed states with variable populations and coherence terms subject to positivity constraints. This motivates an investigation of how acceleration affects entanglement and coherence in initially mixed detector states and how their evolution depends on the initial-state parameters.

In this work, we investigate the evolution of entanglement and coherence in two Unruh-DeWitt detectors initially prepared in a mixed two-qubit X state. One detector remains inertial and uncoupled from the field, while the other undergoes uniform acceleration and interacts locally with a massless scalar field in the Minkowski vacuum. We derive the reduced detector state within a perturbative treatment and quantify its quantum resources using concurrence, the $l_1$ norm of coherence, and the relative entropy of coherence. For the initial-state families considered, the resulting expressions exhibit entanglement sudden death at finite acceleration, whereas both coherence measures remain nonzero beyond the entanglement threshold. Within these families, larger initial entanglement is associated with a higher acceleration threshold for its disappearance. We further examine how the detector energy gap and interaction duration affect the persistence of these resources. These results extend the comparison of entanglement and coherence to initially mixed detector states and clarify how their distinct dependence on populations and off-diagonal correlations shapes their response to accelerated motion within the adopted description.

The remainder of this paper is organized as follows. Section II introduces the Unruh-DeWitt detector model and derives the reduced state of two detectors initially prepared in a mixed two-qubit X state, with one detector undergoing uniform acceleration and coupling locally to a scalar field. Section III examines the evolution of entanglement and coherence as functions of acceleration, detector energy gap, and interaction duration. Section IV summarizes our findings.

\section{Evolution of the detectors' X states under relativistic motion}
We first describe the detector-field interaction in the computational basis and then use the resulting expressions to construct the evolution of an initial mixed X state. Consider two observers, Charlie and David, each carrying a two-level Unruh-DeWitt detector with energy gap $\Omega$. Charlie's detector remains at rest in an inertial frame and is kept switched off, whereas David's detector undergoes uniform acceleration $a$ and interacts with a massless scalar field for a duration $\Delta$. The worldline of David's detector is
\[
t(\tau)=a^{-1}\sinh(a\tau),\qquad x(\tau)=a^{-1}\cosh(a\tau),\qquad y(\tau)=z(\tau)=0,
\]
where $\tau$ is the detector's proper time \cite{SDF53}. Throughout this paper, we use natural units $c=\hbar=k_B=1$. To establish the basis-state evolution, we begin with the following product state of the detector pair and the field:
\begin{equation}
\label{qq1}\left| \Psi_{t_0}^{CD\phi} \right\rangle = \left| \Psi_{CD} \right\rangle \otimes \left| 0_M \right\rangle,
\end{equation}
where $|0_M\rangle$ denotes the Minkowski vacuum of the scalar field. The detector Hilbert space is spanned by $\{|0_C0_D\rangle,|0_C1_D\rangle,|1_C0_D\rangle,|1_C1_D\rangle\}$, with $|0\rangle$ and $|1\rangle$ denoting the ground and excited states, respectively \cite{SDF54}. We first take $|\Psi_{CD}\rangle=|0_C1_D\rangle$ as an illustrative basis state. This intermediate choice is used to derive the basis-state transformations and does not restrict the mixed initial states considered below.

The total Hamiltonian of the system is given by
\begin{eqnarray}\label{A1}
H_{CD\phi}=H_{C}+H_{D}+H_{KG}+H_{int}^{D\phi},
\end{eqnarray}
where $H_C=\Omega C^{\rm\dagger}C$ and $H_D=\Omega D^{\rm\dagger}D$ are the free detector Hamiltonians, and $H_{KG}$ is the free Hamiltonian of the massless scalar field. The operators $C$ and $D$ lower the corresponding detector states. Only David's detector couples to the field, through the interaction Hamiltonian
\begin{eqnarray}\label{A2}
H_{int}^{D\phi}(t)=\epsilon(t)\int_{\sum_{t}}d^{3} \boldsymbol{x}\sqrt{-g}\phi(x)[\psi(\boldsymbol{x})D+\bar{\psi}(\boldsymbol{x})D^{\rm \dagger}],
\end{eqnarray}
where $g=\det(g_{ab})$ and $g_{ab}$ is the spacetime metric in the chosen coordinates. The real switching function $\epsilon(t)$ controls the interaction, while the spatial profile
$\psi(\boldsymbol{x})=(\kappa\sqrt{2\pi})^{-3}\exp[-\boldsymbol{x}^2/(2\kappa^2)]$
localizes the coupling around the detector. The parameter $\kappa$ characterizes the spatial width of this Gaussian profile, which decays rapidly away from the detector center \cite{SDF59}.

In the weak-coupling regime, we expand the interaction-picture evolution to first order in the coupling. At the end of the interaction, denoted by $t=t_0+\Delta$, the detector-field state is approximated by
\begin{eqnarray}\label{qq4}
|\Psi^{CD\phi}_t\rangle=(I+a^{\rm \dagger}_{RI}(\lambda)D-a_{RI}(\bar{\lambda})D^{\rm \dagger})\left| \Psi_{t_0}^{CD\phi} \right\rangle,
\end{eqnarray}
where $\lambda=-KEf$ and $f=\epsilon(t)e^{-i\Omega t}\psi(\boldsymbol{x})$. Here, $E$ is the difference between the advanced and retarded Green functions of the Klein-Gordon equation, and $K$ selects the positive-frequency part with respect to Rindler time. The operators $a_{RI}^{\rm\dagger}$ and $a_{RI}$ create and annihilate field excitations in Rindler region I, respectively \cite{SDF55}.

For the initial detector state $|0_C1_D\rangle$, Eq.~(\ref{qq4}) becomes
\begin{eqnarray}\label{qq5}
|\Psi^{CD\phi}_t\rangle=(I+a^{\rm \dagger}_{RI}(\lambda)D-a_{RI}(\bar{\lambda})D^{\rm \dagger})|0_C1_D\rangle\otimes |0_M\rangle.
\end{eqnarray}
To evaluate the action of the Rindler operators on the Minkowski vacuum, we use the Bogoliubov relations
\begin{eqnarray}\label{qq6}
a_{RI}(\bar{\lambda})=\frac{a_{M}(\overline{F_{1\Omega}})+e^{-\pi\Omega/a}a^{\rm \dagger}_{M}(F_{2\Omega})}{(1-e^{-2\pi\Omega/a})^{1/2}},
\end{eqnarray}
\begin{eqnarray}\label{qq7}
a^{\rm \dagger}_{RI}(\lambda)=\frac{a^{\rm \dagger}_{M}(F_{1\Omega})+e^{-\pi\Omega/a}a_{M}(\overline{F_{2\Omega}})}{(1-e^{-2\pi\Omega/a})^{1/2}},
\end{eqnarray}
where $F_{1\Omega}=\frac{\lambda+e^{-\pi\Omega/a}\lambda\circ w}{(1-e^{-2\pi\Omega/a})^{1/2}}$ and $F_{2\Omega}=\frac{\overline{\lambda\circ w}+e^{-\pi\Omega/a}\bar{\lambda}}{(1-e^{-2\pi\Omega/a})^{1/2}}$. The wedge reflection $w(t,x,y,z)=(-t,-x,y,z)$ relates Rindler regions I and II, and $\lambda\circ w$ denotes the pullback of $\lambda$ under this map. Substituting Eqs.~(\ref{qq6}) and (\ref{qq7}) into Eq.~(\ref{qq5}) gives
\begin{eqnarray}\label{qq8}
|\Psi^{CD\phi}_{t|01\rangle}\rangle=|01\rangle\otimes |0_M\rangle+\nu\frac{|00\rangle \otimes |1_{\tilde{F}_{1\Omega}}\rangle}{\left(1 - e^{-2\pi \Omega/a}\right)^{\frac{1}{2}}},
\end{eqnarray}
where $\widetilde{F}_{i\Omega}=F_{i\Omega}/\nu$, and $|1_{\widetilde F_{i\Omega}}\rangle$ denotes the corresponding one-particle field state. Applying the same procedure to the other computational basis states gives the following states in $\mathcal H_C\otimes\mathcal H_D\otimes\mathcal H_\phi$:
\begin{eqnarray}\label{qq9}
|\Psi^{CD\phi}_{t|00\rangle}\rangle=|00\rangle\otimes |0_M\rangle+\nu e^{-\pi \Omega/a} \frac{|01\rangle \otimes |1_{\tilde{F}_{2\Omega}}\rangle}{\left(1 - e^{-2\pi \Omega/a}\right)^{\frac{1}{2}}},
\end{eqnarray}
\begin{eqnarray}\label{qq10}
|\Psi^{CD\phi}_{t|10\rangle}\rangle=|10\rangle\otimes |0_M\rangle+\nu e^{-\pi \Omega/a} \frac{|11\rangle \otimes |1_{\tilde{F}_{2\Omega}}\rangle}{\left(1 - e^{-2\pi \Omega/a}\right)^{\frac{1}{2}}},
\end{eqnarray}
\begin{eqnarray}\label{qq11}
|\Psi^{CD\phi}_{t|11\rangle}\rangle=|11\rangle\otimes |0_M\rangle+\nu\frac{|10\rangle \otimes |1_{\tilde{F}_{1\Omega}}\rangle}{\left(1 - e^{-2\pi \Omega/a}\right)^{\frac{1}{2}}}.
\end{eqnarray}

The reduced detector operators are obtained by tracing over the external field. For the diagonal and antidiagonal basis operators needed to construct an X state, the resulting expressions are
\begin{align*}
{\rm{tr}}_{\phi}|\Psi^{CD\phi}_{t|00\rangle}\rangle\langle \Psi^{CD\phi}_{t|00\rangle}| &= |00\rangle\langle 00|+\frac{\nu^{2}e^{-2\pi \Omega/a}}{1 - e^{-2\pi \Omega/a}}|01\rangle\langle 01|,\quad {\rm{tr}}_{\phi}|\Psi^{CD\phi}_{t|00\rangle}\rangle\langle \Psi^{CD\phi}_{t|11\rangle}|=|00\rangle\langle 11|, \\
{\rm{tr}}_{\phi}|\Psi^{CD\phi}_{t|01\rangle}\rangle\langle \Psi^{CD\phi}_{t|01\rangle}| &= |01\rangle\langle 01|+\frac{\nu^{2}}{1 - e^{-2\pi \Omega/a}}|00\rangle\langle 00|,\quad {\rm{tr}}_{\phi}|\Psi^{CD\phi}_{t|11\rangle}\rangle\langle \Psi^{CD\phi}_{t|00\rangle}|=|11\rangle\langle 00|, \\
{\rm{tr}}_{\phi}|\Psi^{CD\phi}_{t|10\rangle}\rangle\langle \Psi^{CD\phi}_{t|10\rangle}| &= |10\rangle\langle 10|+\frac{\nu^{2}e^{-2\pi \Omega/a}}{1 - e^{-2\pi \Omega/a}}|11\rangle\langle 11|,\quad {\rm{tr}}_{\phi}|\Psi^{CD\phi}_{t|01\rangle}\rangle\langle \Psi^{CD\phi}_{t|10\rangle}|=|01\rangle\langle 10|, \\
{\rm{tr}}_{\phi}|\Psi^{CD\phi}_{t|11\rangle}\rangle\langle \Psi^{CD\phi}_{t|11\rangle}| &= |11\rangle\langle 11|+\frac{\nu^{2}}{1 - e^{-2\pi \Omega/a}}|10\rangle\langle 10|,\quad {\rm{tr}}_{\phi}|\Psi^{CD\phi}_{t|10\rangle}\rangle\langle \Psi^{CD\phi}_{t|01\rangle}|=|10\rangle\langle 01|.
\end{align*}

We now apply these basis-state expressions to an initial mixed detector state. For this part of the analysis, we denote the inertial detector by Alice ($A$) and the accelerated detector by Bob ($B$); their roles are identical to those of Charlie and David above. We assume that the initial total state is $\rho^X_{AB}\otimes|0_M\rangle\langle0_M|$. In the ordered computational basis $\{|00\rangle,|01\rangle,|10\rangle,|11\rangle\}$, a general two-qubit X state has the form
\begin{equation}\label{qq16}
    \rho^{X}_{AB} =
    \begin{pmatrix}
        \rho_{11} & 0 & 0 & \rho_{14} \\
        0 & \rho_{22} & \rho_{23} & 0 \\
        0 & \rho_{32} & \rho_{33} & 0 \\
        \rho_{41} & 0 & 0 & \rho_{44}
    \end{pmatrix}.
\end{equation}
Hermiticity requires $\rho_{41}=\rho_{14}^*$ and $\rho_{32}=\rho_{23}^*$. The diagonal elements are nonnegative and satisfy $\rho_{11}+\rho_{22}+\rho_{33}+\rho_{44}=1$, while positivity further requires $\rho_{22}\rho_{33}\geq|\rho_{23}|^2$ and $\rho_{11}\rho_{44}\geq|\rho_{14}|^2$. The state is entangled if $|\rho_{14}|^2>\rho_{22}\rho_{33}$ or $|\rho_{23}|^2>\rho_{11}\rho_{44}$.

Combining the field-traced basis operators with the initial matrix elements and normalizing the resulting expression gives the reduced detector state within the adopted first-order treatment of the state vector. This expression retains the X structure:
\begin{equation}\label{qq17}
    \rho^{Xt}_{AB} =
    \begin{pmatrix}
        \rho_{11}\alpha+\rho_{22}\beta & 0 & 0 & \rho_{14}\alpha \\
        0 & \rho_{22}\alpha+\rho_{11}\gamma & \rho_{23}\alpha & 0 \\
        0 & \rho_{32}\alpha & \rho_{33}\alpha+\rho_{44}\beta & 0 \\
        \rho_{41}\alpha & 0 & 0 & \rho_{44}\alpha+\rho_{33}\gamma
    \end{pmatrix}.
\end{equation}
Here, $\rho_{ij}$ continue to denote the initial matrix elements, and the coefficients $\alpha$, $\beta$, and $\gamma$ are
\[
   \alpha=\frac{1-q}{(\rho_{11}+\rho_{22}+\rho_{33}+\rho_{44})(1-q)+\nu^{2}(\rho_{22}+\rho_{44}+q(\rho_{11}+\rho_{33}))},
\]
\[
   \beta=\frac{\nu^{2}}{(\rho_{11}+\rho_{22}+\rho_{33}+\rho_{44})(1-q)+\nu^{2}(\rho_{22}+\rho_{44}+q(\rho_{11}+\rho_{33}))},
\]
\[
   \gamma=\frac{\nu^{2}q}{(\rho_{11}+\rho_{22}+\rho_{33}+\rho_{44})(1-q)+\nu^{2}(\rho_{22}+\rho_{44}+q(\rho_{11}+\rho_{33}))}.
\]
We use the dimensionless acceleration parameter $q=e^{-2\pi\Omega/a}$ and the effective coupling parameter
$\nu^2\equiv\|\lambda\|^2=\frac{\epsilon^2\Omega\Delta}{2\pi}e^{-\Omega^2\kappa^2}$.
The latter expression assumes $\Omega^{-1}\ll\Delta$, and weak coupling requires $\nu^2\ll1$. At fixed $\Omega$, $q$ increases monotonically with $a$, approaching zero in the zero-acceleration limit and unity in the infinite-acceleration limit. The transition weights also contain $\nu^2/(1-q)$, so weak coupling alone does not justify extrapolating the perturbative description to arbitrarily large acceleration. In the following section, we use $\rho^{Xt}_{AB}$ to examine entanglement and coherence within this description.

\section{Unruh effect on quantum  entanglement and coherence of generic X-states}
We now use the reduced detector state $\rho^{Xt}_{AB}$ to compare the evolution of entanglement and coherence under accelerated motion. We first examine their dependence on the acceleration parameter $q$ for several initial X states and then consider the detector energy gap and interaction duration. Entanglement is quantified by the concurrence \cite{SDF61,SDF62},
\begin{equation}\label{qq18}
    E(\rho^{Xt}_{AB}) = \max\{0, \sqrt{\lambda_1} - \sqrt{\lambda_2} - \sqrt{\lambda_3} - \sqrt{\lambda_4}\},\quad\lambda_i \ge \lambda_{i+1} \ge 0,
\end{equation}
where $\lambda_i$ are the eigenvalues of $\rho^{Xt}_{AB}\widetilde{\rho^{Xt}_{AB}}$ in decreasing order, with $\widetilde{\rho^{Xt}_{AB}}=(\sigma_y\otimes\sigma_y)(\rho^{Xt}_{AB})^*(\sigma_y\otimes\sigma_y)$ denoting the spin-flipped state. Here, complex conjugation is taken in the computational basis, and $\sigma_y$ is the Pauli-$y$ matrix.
Substituting the evolved density matrix in Eq.~(\ref{qq17}) into Eq.~(\ref{qq18}), we obtain
\begin{equation}\label{qq19}
\begin{aligned}
E(\rho^{Xt}_{AB})=2\max\Bigl\{0,\;&
\alpha|\rho_{14}|-
\sqrt{(\alpha\rho_{22}+\gamma\rho_{11})(\alpha\rho_{33}+\beta\rho_{44})},\\
&\alpha|\rho_{23}|-
\sqrt{(\alpha\rho_{11}+\beta\rho_{22})(\alpha\rho_{44}+\gamma\rho_{33})}
\Bigr\}.
\end{aligned}
\end{equation}
Here, $\rho_{ij}$ are the initial matrix elements in Eq.~(\ref{qq16}), and $\alpha$, $\beta$, and $\gamma$ are the coefficients in Eq.~(\ref{qq17}). Each of the two competing terms in the concurrence compares the magnitude of an evolved off-diagonal element with a threshold determined by the diagonal populations. Entanglement disappears when neither off-diagonal magnitude exceeds its corresponding threshold.

\begin{figure}
\begin{minipage}[t]{0.5\linewidth}
\centering
\includegraphics[width=3.0in,height=5.2cm]{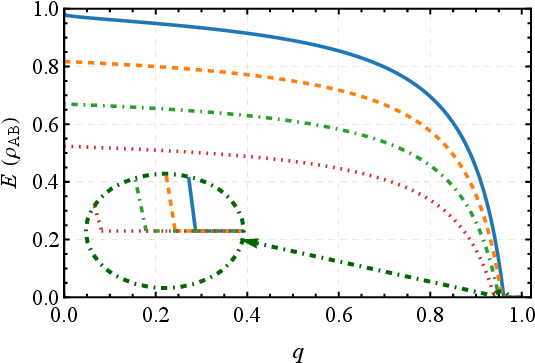}
\label{fig1a}
\end{minipage}%
\begin{minipage}[t]{0.5\linewidth}
\centering
\label{fig1b}
\end{minipage}%
\caption{Concurrence $E(\rho^{Xt}_{AB})$ as a function of the acceleration parameter $q$ at fixed $\nu^2=0.04$. From top to bottom, the four curves correspond to the following initial-state parameters: (i) $\rho_{11}=\rho_{44}=\rho_{14}=\rho_{41}=0.5$ and $\rho_{22}=\rho_{33}=\rho_{23}=\rho_{32}=0$; (ii) $\rho_{11}=\rho_{44}=\dfrac{19}{40}$, $\rho_{22}=\rho_{33}=\dfrac{1}{40}$, $\rho_{14}=\rho_{41}=\dfrac{9}{20}$, and $\rho_{23}=\rho_{32}=0$; (iii) $\rho_{11}=\rho_{44}=\dfrac{9}{20}$, $\rho_{22}=\rho_{33}=\dfrac{1}{20}$, $\rho_{14}=\rho_{41}=\dfrac{2}{5}$, and $\rho_{23}=\rho_{32}=0$; and (iv) $\rho_{11}=\rho_{44}=\dfrac{17}{40}$, $\rho_{22}=\rho_{33}=\dfrac{3}{40}$, $\rho_{14}=\rho_{41}=\dfrac{7}{20}$, and $\rho_{23}=\rho_{32}=0$.}
\label{Fig1}
\end{figure}

Fig.~\ref{Fig1} shows the concurrence $E(\rho^{Xt}_{AB})$ as a function of the acceleration parameter $q$ at a fixed effective coupling of $\nu^2=0.04$. For the selected initial states, the concurrence decreases with increasing $q$ and vanishes at a finite value of this parameter. The inset further shows that, within this family, larger initial entanglement corresponds to a higher threshold for its disappearance. This behavior arises because the interaction between the accelerated detector and the field modifies both the detector populations and the off-diagonal correlations: the off-diagonal elements are scaled by $\alpha$, while the populations also receive transition contributions proportional to $\beta$ and $\gamma$. When the off-diagonal correlations no longer exceed the population-dependent thresholds in Eq.~(\ref{qq19}), the detector state crosses the boundary between entangled and separable states. Entanglement sudden death therefore does not require all coherence terms to vanish; instead, it reflects the threshold behavior associated with crossing the separability boundary during continuous state evolution. Within the selected family, stronger initial entanglement provides a larger margin against acceleration-induced decoherence, allowing entanglement to persist to higher values of the acceleration parameter. This result offers insight into the use of entanglement for relativistic quantum information processing. For quantum teleportation, entanglement-assisted communication, and distributed quantum tasks, initial-state preparation should account for both the amount of entanglement and the associated population and coherence structure to improve the persistence of the resource under noninertial conditions. The threshold for entanglement disappearance also provides a reference for assessing the parameter range over which shared entanglement remains available, thereby informing initial-state selection and the choice of operating parameters.

We next quantify coherence in the fixed computational basis $\{|00\rangle,|01\rangle,|10\rangle,|11\rangle\}$ using the $l_1$ norm and the relative entropy of coherence (REC) \cite{zxl12}. These measures provide complementary descriptions of the off-diagonal structure of $\rho^{Xt}_{AB}$. The $l_1$ norm is
\begin{equation}\label{qq20}
   C_{l_1}(\rho^{Xt}_{AB}) = \sum_{\substack{i,j=1\\i\neq j}}^4 |\tilde{\rho}_{ij}|,
\end{equation}
where $\tilde{\rho}_{ij}=\langle i|\rho^{Xt}_{AB}|j\rangle$, with $i,j=1,\ldots,4$ labeling the ordered computational basis states. We retain $\rho_{ij}$ for the initial matrix elements. Using $\rho_{41}=\rho_{14}^*$ and $\rho_{32}=\rho_{23}^*$, Eq.~(\ref{qq17}) gives
\begin{equation}\label{qq21}
   C_{l_1}(\rho^{Xt}_{AB}) = 2\alpha\left(|\rho_{14}|+|\rho_{23}|\right).
\end{equation}
Here, $\alpha$ is real and nonnegative, and the magnitudes of the two independent off-diagonal elements contribute separately. The REC is defined by
\begin{equation}\label{qq22}
   C_{\mathrm{REC}}(\rho^{Xt}_{AB}) = S((\rho^{Xt}_{AB})_{\mathrm{diag}}) - S(\rho^{Xt}_{AB}),
\end{equation}
where $S$ is the von Neumann entropy and $(\rho^{Xt}_{AB})_{\mathrm{diag}}$ denotes the quantum state obtained by setting all off-diagonal elements of $\rho^{Xt}_{AB}$ to zero. Expressing the entropies in terms of the diagonal elements and eigenvalues of the evolved state gives
\begin{equation}\label{qq23}
    C_{\mathrm{REC}}(\rho^{Xt}_{AB}) = -\sum_{i=1}^4 \tilde{\rho}_{ii} \log_2 \tilde{\rho}_{ii} + \sum_{k=1}^4 \xi_k \log_2 \xi_k,
\end{equation}
where the eigenvalues are
\[
  \xi_1 = \frac{1}{2} \left[ (\tilde{\rho}_{11} + \tilde{\rho}_{44}) - \sqrt{\tilde{\rho}_{11}^2+4\tilde{\rho}_{14}\tilde{\rho}_{41}-2\tilde{\rho}_{11}\tilde{\rho}_{44} + \tilde{\rho}_{44}^2} \right],
\]
\[
  \xi_2 = \frac{1}{2} \left[ (\tilde{\rho}_{11} + \tilde{\rho}_{44}) + \sqrt{\tilde{\rho}_{11}^2+4\tilde{\rho}_{14}\tilde{\rho}_{41}-2\tilde{\rho}_{11}\tilde{\rho}_{44} + \tilde{\rho}_{44}^2} \right],
\]
\[
  \xi_3 = \frac{1}{2} \left[ (\tilde{\rho}_{22} + \tilde{\rho}_{33}) - \sqrt{\tilde{\rho}_{22}^2+4\tilde{\rho}_{23}\tilde{\rho}_{32}-2\tilde{\rho}_{22}\tilde{\rho}_{33} + \tilde{\rho}_{33}^2} \right],
\]
\[
  \xi_4 = \frac{1}{2} \left[ (\tilde{\rho}_{22} + \tilde{\rho}_{33}) + \sqrt{\tilde{\rho}_{22}^2+4\tilde{\rho}_{23}\tilde{\rho}_{32}-2\tilde{\rho}_{22}\tilde{\rho}_{33} + \tilde{\rho}_{33}^2} \right].
\]
Here, $\tilde{\rho}_{ij}$ are the evolved matrix elements defined above, and $\xi_k$ ($k=1,\ldots,4$) are the eigenvalues of $\rho^{Xt}_{AB}$. We use the convention $0\log_2 0=0$.

\begin{figure}
\begin{minipage}[t]{0.5\linewidth}
\centering
\includegraphics[width=3.0in,height=5.2cm]{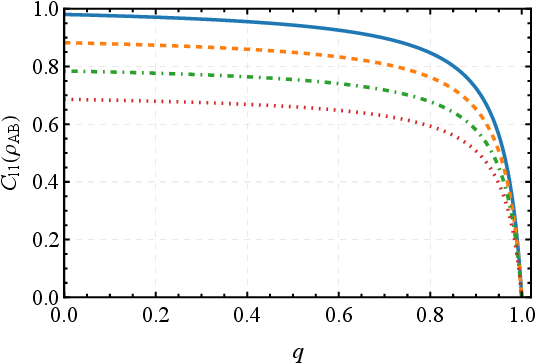}
\label{pic2}
\end{minipage}%
\begin{minipage}[t]{0.5\linewidth}
\centering
\includegraphics[width=3.0in,height=5.2cm]{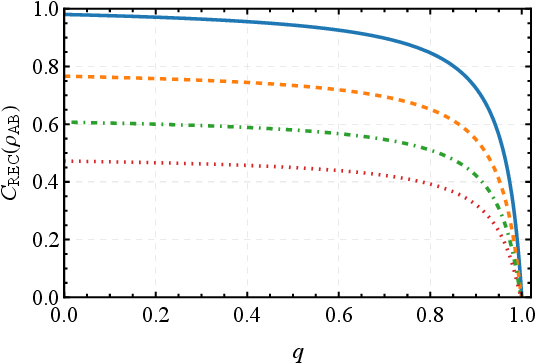}
\label{pic3}
\end{minipage}%
\caption{Evolution of the $l_1$ norm of coherence $C_{l_{1}}(\rho^{Xt}_{AB})$ (left panel) and the REC $C_{\mathrm{REC}}(\rho^{Xt}_{AB})$ (right panel) as functions of the acceleration parameter $q$. The effective coupling parameter is set to $\nu^{2}=0.04$. The initial X-state parameters for the four curves, from top to bottom, are the same as those used in Fig.~\ref{Fig1}.}
\label{Fig2}
\end{figure}

Fig.~\ref{Fig2} displays the $l_1$ norm of  coherence and REC for the same initial configurations and coupling parameter as Fig.~\ref{Fig1}. Both measures decrease monotonically with $q$ for the states shown, while the configuration with larger initial coherence retains a larger residual value throughout the plotted range. For the $l_1$ norm, this behavior follows directly from Eq.~(\ref{qq21}), in which the acceleration dependence enters through the common coefficient $\alpha$. The REC also depends on the evolved populations through the entropy difference in Eq.~(\ref{qq22}), but exhibits the same qualitative decay for these configurations. Comparing Figs.~\ref{Fig1} and \ref{Fig2} clarifies the origin of their different persistence properties: entanglement vanishes when neither off-diagonal correlation exceeds its population-dependent threshold, whereas coherence remains nonzero whenever the evolved state has a nonzero off-diagonal element in the chosen basis. At fixed $\nu>0$, the coefficient $\alpha$ in Eq.~(\ref{qq17}) is positive for $0\leq q<1$. An initially coherent X state therefore retains nonzero $l_1$ norm of  coherence throughout this interval, and its REC is also nonzero because the state remains nondiagonal. In the formal limit $q\to1$ at fixed nonzero coupling, $\alpha\to0$, the evolved state becomes diagonal, and both coherence measures approach zero. Thus, within the accelerated-detector model considered here, coherence can persist after entanglement has disappeared. This persistence motivates the exploration of quantum information protocols that use coherence in noninertial settings and suggests that the survival of coherence should be considered when selecting resources and operating parameters for such protocols.

\begin{figure*}[t!] 
\centering

\begin{minipage}[t]{0.32\linewidth}
\centering
\includegraphics[width=\linewidth]{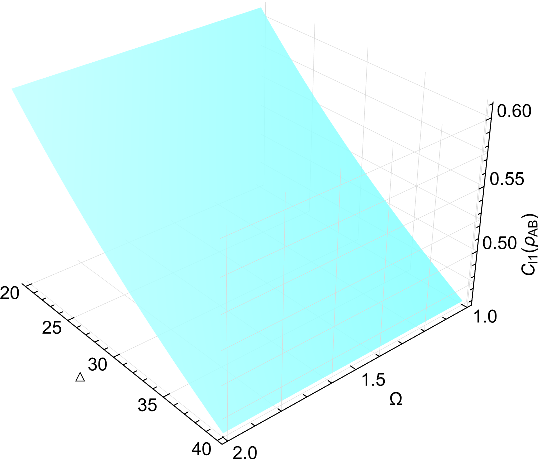}
\vspace{0.1cm}
\centerline{(a)}
\label{zxla}
\end{minipage}\hfill
\begin{minipage}[t]{0.32\linewidth}
\centering
\includegraphics[width=\linewidth]{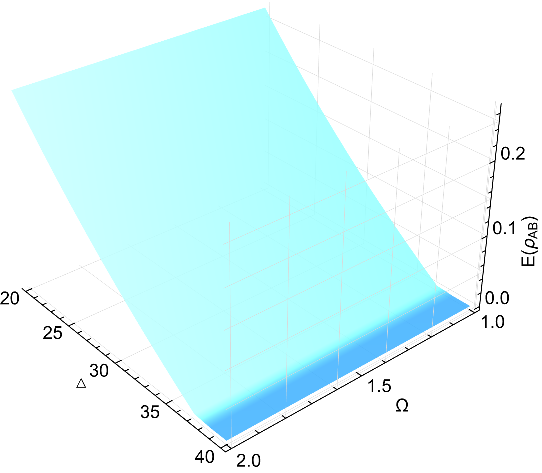}
\vspace{0.1cm}
\centerline{(b)}
\label{zxlb}
\end{minipage}\hfill
\begin{minipage}[t]{0.32\linewidth}
\centering
\includegraphics[width=\linewidth]{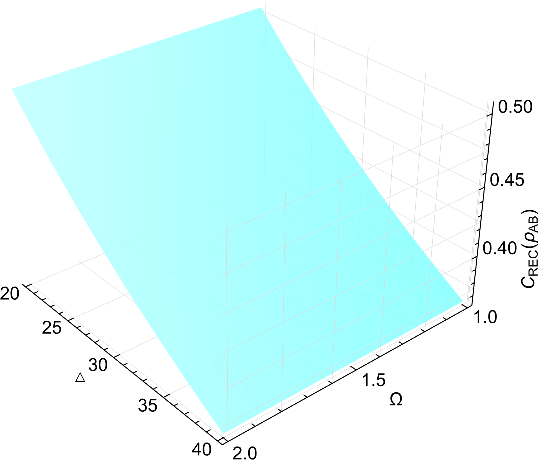}
\vspace{0.1cm}
\centerline{(c)}
\label{zxlc}
\end{minipage}

\vspace{-0.8cm} 

\begin{minipage}[t]{0.32\linewidth}
\centering
\includegraphics[width=\linewidth]{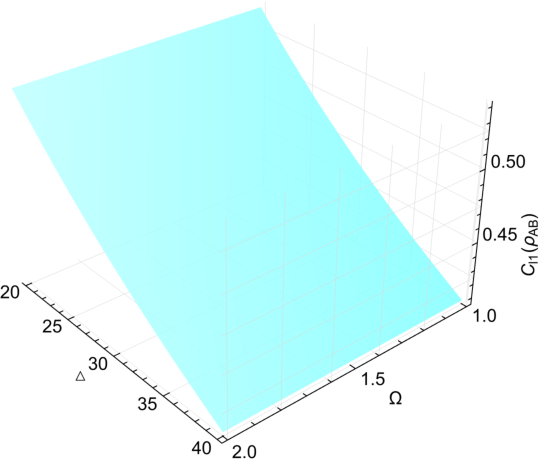}
\vspace{0.1cm}
\centerline{(d)}
\label{zxld}
\end{minipage}\hfill
\begin{minipage}[t]{0.32\linewidth}
\centering
\includegraphics[width=\linewidth]{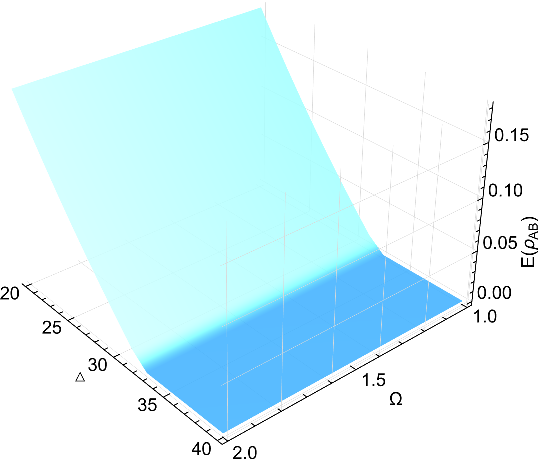}
\vspace{0.1cm}
\centerline{(e)}
\label{zxle}
\end{minipage}\hfill
\begin{minipage}[t]{0.32\linewidth}
\centering
\includegraphics[width=\linewidth]{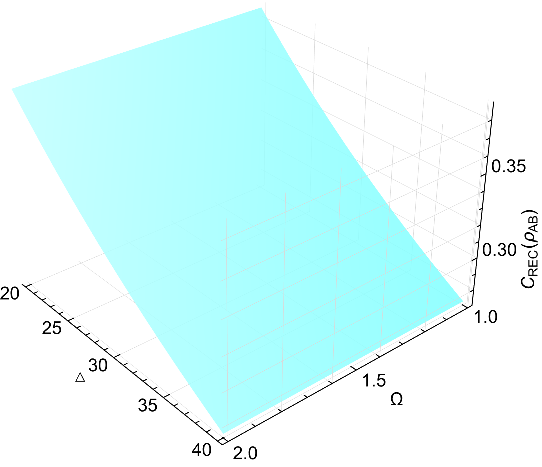}
\vspace{0.1cm}
\centerline{(f)}
\label{zxlf}
\end{minipage}

\caption{The $l_1$ norm of coherence $C_{l_1}(\rho_{AB}^{Xt})$, concurrence $E(\rho_{AB}^{Xt})$, and REC $C_{\mathrm{REC}}(\rho_{AB}^{Xt})$ of the evolved detector state as functions of the detector energy gap $\Omega$ and interaction duration $\Delta$ at fixed acceleration $a=12000$, with $q=\exp(-2\pi\Omega/a)$, $\epsilon^{2}=8\pi^{2}\times10^{-6}$, and $\kappa=10^{-4}$. The left, middle, and right columns show $C_{l_1}(\rho_{AB}^{Xt})$, $E(\rho_{AB}^{Xt})$, and $C_{\mathrm{REC}}(\rho_{AB}^{Xt})$, respectively. The initial X states for the upper row [(a)--(c)] and lower row [(d)--(f)] correspond to the second and third configurations in Fig.~\ref{Fig1}, respectively.}
\label{Fig3}
\end{figure*}

Fig.~3 further examines the dependence of quantum resources on the detector energy gap $\Omega$ and interaction duration $\Delta$ at fixed acceleration $a=12000$, with $q=\exp(-2\pi\Omega/a)$, $\epsilon^2=8\pi^2\times10^{-6}$, and $\kappa=10^{-4}$.  The upper and lower rows correspond to the second and third initial configurations in Fig.~\ref{Fig1}, respectively, and the panels show the $l_1$ norm of coherence, concurrence, and REC. Over the displayed parameter range, the concurrence surfaces contain extended regions of zero entanglement, whereas both coherence measures remain nonzero and vary smoothly over the corresponding regions. This distinction arises from the different conditions under which the two resources are nonzero: entanglement requires off-diagonal correlations to exceed population-dependent thresholds, whereas coherence can persist when this condition is no longer satisfied. Comparing the two rows further shows that, within the selected initial-state family, the configuration with larger initial entanglement and coherence retains larger residual resource values. These results indicate that the detector-field interaction plays an important role in determining the persistence of quantum resources under relativistic motion by modifying the detector populations and off-diagonal correlations. Detector configuration and initial-state structure should therefore be considered jointly when designing relativistic quantum information protocols. In particular, coherence that survives after entanglement disappears motivates the exploration of quantum information protocols that use coherence in noninertial settings. From a theoretical perspective, the dependence of resource persistence on energy gap and interaction duration may inform the design of detection schemes based on artificial two-level systems with tunable energy gaps and suggests that detector parameters and the initial mixed-state structure can be selected jointly to improve resource persistence. When combined with a dynamical description valid in the target parameter regime and a specific implementation platform, such parameter-selection strategies could provide a basis for developing experimentally realizable schemes for detection and quantum information processing in relativistic settings.

\section{Conclusions}
We have investigated the evolution of entanglement and coherence in two Unruh-DeWitt detectors initially prepared in a mixed two-qubit X state. One detector remains inertial and uncoupled from the field, while the other undergoes uniform acceleration and interacts locally with a massless scalar field in the Minkowski vacuum. By evaluating the concurrence, the $l_1$ norm of coherence, and the REC of the evolved detector state, we have compared the responses of these resources to accelerated motion. For the initial-state families considered, the resulting expressions show that entanglement decreases with the acceleration parameter and vanishes at a finite threshold, whereas both coherence measures remain nonzero after entanglement has disappeared. Within these families, larger initial entanglement corresponds to a higher threshold for its disappearance. We have also examined the influence of the detector energy gap and interaction duration on resource persistence, further illustrating the distinct evolution of entanglement and coherence.

This distinction arises from the different conditions for the presence of the two resources: entanglement requires off-diagonal correlations to exceed population-dependent thresholds, whereas coherence requires only nonzero off-diagonal elements in a specified reference basis. The disappearance of entanglement therefore does not imply the simultaneous loss of detector coherence. For experimentally prepared states, which generally possess some degree of mixedness, these results suggest that resource stability should be assessed by considering both the initial resource content and the underlying population and coherence structure. For relativistic quantum information tasks that rely on shared entanglement, initial-state selection and the parameter range over which entanglement survives are important considerations in protocol design. At the same time, coherence that persists after entanglement disappears motivates the exploration of quantum information protocols that use coherence in noninertial settings. Future work could employ dynamical methods applicable to longer interaction durations or higher accelerations to test these persistence features and extend the analysis to specific gravitational backgrounds, thereby exploring the interplay between initial mixedness, detector response, and quantum resource evolution.

\begin{acknowledgments}
This work is supported by the National Natural Science Foundation of China (12575056), the Natural
Science Foundation of the Liaoning Scientific Committee (2026-MS-281), the LiaoNing Revitalization Talents Program (XLYC2503099), the Joint Funds Program of Liaoning Provincial Natural Science Foundation (2025-BSLH-223), and the Technology Talent Innovation Support Policy Project Plan of Dalian (2024RQ051).
\end{acknowledgments}


\end{document}